\documentclass{aa}
\usepackage{apj2aa}
\usepackage{astron} 
\usepackage{float,epsfig,psfig}
\usepackage{pstricks}

\def\hea4{{\it HEAO~A4}}
\def\heaoa2{{\it HEAO~A2}}
\def\heao1{{\it HEAO~1}}

\def\h0{$H_{\rm o}=50$~km~s$^{-1}$~Mpc$^{-1}$}
\def\q0{$q_{\rm o}$}

\def\msun     {M$_{\odot}$}

\def\etal    {{et~al.}~}

\def\cms3  {~{cm$^{-3}$}}

\newcommand{\mincir}{\raise
  -2.truept\hbox{\rlap{\hbox{$\sim$}}\raise5.truept \hbox{$<$}\ }}
\newcommand{\magcir}{\raise
  -2.truept\hbox{\rlap{\hbox{$\sim$}}\raise5.truept \hbox{$>$}\ }}
\newcommand{\hm}{\,h^{-1}{\rm Mpc}}

\begin{document}

\title{Reproducing  the entropy structure in galaxy groups}
\author{A. Finoguenov\inst{1}, S. Borgani\inst{2,3},
L. Tornatore\inst{2}, H. B\"ohringer\inst{1}}

\institute{Max-Planck Institut f\"ur extraterrestrische Physik,
             Giessenbachstra\ss e 1, D-85748 Garching, Germany 
\and
             Dipartimento di Astronomia, Universit\`a di Trieste, via
             Tiepolo 11, I-34131 Trieste, Italy 
\and 
             INAF, Istituto
             Nazionale di Astrofisica; INFN, Sezione di Trieste,
             Trieste, Italy }

\date{Received October 23 2002; accepted December 19 2002}
\authorrunning{Finoguenov \etal}

\abstract{We carry out a comparison between observations and hydrodynamic
simulations of entropy profiles of groups and clusters of galaxies.  We use
the Tree+SPH GADGET code to simulate four halos of sizes in the
$M_{500}=1.0-16\times10^{13}h^{-1}M_{\odot}$\ range, corresponding to poor
groups up to Virgo-like clusters. We concentrate on the effect of
introducing radiative cooling, star formation, and a variety of
non-gravitational heating schemes on the entropy structure and the stellar
fraction. We show that all the simulations result in a correct entropy
profile for the Virgo-like cluster. With the heating energy budget of $\sim
0.7$ keV/particle injected at $z_h=3$, we are also able to reproduce the
entropy profiles of groups. We obtain the flat entropy cores as a combined
effect of preheating and cooling, while we achieve the high entropy at
outskirts by preheating.  The resulting baryon fraction locked into stars is
in the 25--30\% range, compared to 35--40\% in the case of no
preheating. Heating at higher redshift, $z_h=9$, strongly delays the
star-formation, but fails to produce a sufficiently high specific
entropy. }

\maketitle

\section{Introduction}
Clusters of galaxies represent invaluable astrophysical laboratories
for the study of the evolution of diffuse cosmic baryons and their
interplay with the processes of star formation and galaxy
evolution. The observational determination of scaling relations
between X-ray properties, such as luminosity $L_X$, gas temperature
$T$, and entropy have now demonstrated that this interplay is crucial
in establishing the physical properties of the intra-cluster medium
(ICM). The slope of the $L_X-T$ relation (e.g.  Markevitch 1998), the
amplitude of the mass--temperature relation (e.g. Finoguenov,
Reiprich, B\"ohringer 2001b) and the gas entropy level (e.g. Ponman,
Cannon \& Navarro 1998; Finoguenov et al. 2002) are all at variance
with respect to model predictions based on pure gravitational heating
(Kaiser 1986) and call for the need of introducing extra physics to
describe the thermodynamics of the ICM (e.g. Evrard \& Henry 1991).
The influence of the energy feedback on the resulting ICM properties caused
in turn an interest in using clusters of galaxies as fossil records of the
past history of star formation (e.g. Menci \& Cavaliere 2000; Bower et
al. 2001) and the corresponding metal production (e.g. Renzini 1997,
Finoguenov, Arnaud \& David 2001a, Pipino et al. 2002). Following the
approach developed in a series of works (e.g. Bower 1997; Ponman, Cannon,
Navarro 1999; Tozzi \& Norman 2001; Borgani et al. 2001), we use the entropy
to characterize the properties of the ICM. We adopt the entropy estimator
$S={kT / n_e^{2/3}}$ [keV cm$^2$], where $T$ is the gas temperature and
$n_e$ is the local electron number density.
So far the reproduction of the cluster properties in simulations has
generally been restricted to a comparison with general scaling properties of
the clusters, such as the $L_X-T$ and the $M-T$ relations (Bialek, Evrard \&
Mohr 2001; Borgani et al. 2002; Muanwong et al. 2002; Dav\'e, Katz \&
Weinberg 2002; Kay, Thomas \& Theuns 2002; and references therein).  In this
{\it Letter} we carry out a detailed comparison between the simulations,
including the effect of radiative cooling and non-gravitational heating, and
X-ray observations of entropy profiles in groups and clusters of
galaxies. Our prime goals are to check under which conditions the shape of
the observed profiles and their scatter can be reproduced and which is the
effect of cooling/heating in determining both the entropy characteristics of
the ICM and the amount of gas locked into the cold (stellar) phase.

\section{Description of simulations and observations}

We use GADGET, a fully parallel Tree+SPH code with adaptive time-stepping
(Springel, Yoshida \& White 2001). The version of the code used here
includes the effect of radiative cooling and photo-ionizing UV background,
while using an entropy-conserving implementation of the SPH with arithmetic
symmetrization of the pairwise pressure force among gas particles (Springel
\& Hernquist 2002).
We simulate four halos, having virial mass in the range $2.3\times 10^{13}
M_{\odot}\mincir M_{\rm vir}\mincir 3.9\times 10^{14}M_{\odot}$, selected from a
DM-only simulation of a $\Lambda$CDM model with $\Omega_m=0.3$, $h=0.7$,
$\sigma_8=0.8$ and $f_{\rm bar}=0.13$ for the baryon fraction, within a box of
size $70\hm$ (see Tornatore et al. 2002, for a detailed description of the
simulations). The simulated structures encompass the mass range
characteristic of moderately rich clusters to groups, where the effect of
non-gravitational gas heating is expected to be more important.
Following the method by Katz \& White (1993), mass and force resolution is
increased in the Lagrangian patch of the simulation which form by $z=0$ a
region encompassing several virial radii of the selected halos. Resolution
is progressively degraded in the outer parts, so as to save computational
time, while providing a correct representation of the large-scale tidal
field.  In the high resolution region, we choose the $\epsilon_{\rm
Pl}=5\,h^{-1}$ kpc for the Plummer-equivalent softening scale, which is kept
fixed in physical scale out to $z=2$, while it is $\epsilon_{\rm
Pl}=15\,h^{-1}$ kpc fixed in co-moving scale at earlier times. Gas particles
in the high-resolution region have a mass $m_{\rm gas}\simeq 3.1\times
10^8M_{\odot}$. Since finite resolution could lead to an underestimate of
the fraction of gas converted into stars (e.g. Balogh et al. 2001; Tornatore
et al. 2002), several runs of groups have been realized at $2^3$ better mass
resolution, $m_{\rm gas}\simeq 3.9\times 10^7M_{\odot}$ and twice as small
force softening scale. Therefore, while the low-resolution group runs only
provide lower limits to the fraction $f_{*}$ of gas locked into stars, the
high-resolution runs are able to resolve galaxy-sized halos well below $L_*$
and provide numerically more reliable results on the stellar mass fraction.
As such, our simulations represent among the highest-resolution attempt to
characterize the gas cooling within groups and clusters of galaxies.
The main characteristics of the simulations are described in Table 1, where
for each structure we give the mass within overdensity
$\Delta_{500}=\rho/\rho_{\rm cr}=500$, $M_{500}$ (units of $10^{13}M_\odot$) and
the number of gas particles, $N_g$, within the virial radius at $z=0$ in the
runs with gravitational heating only.  Besides including only gravitational
gas heating (GH in the notation of Table 1), our runs include the effects of
cooling and the star formation recipe by Katz, Weinberg \& Hernquist (1996),
which converts dense and cold gas particles into collisionless stars (SF
runs). Since no successful approach has been yet developed to include in
simulations the effect of supenova (SN) energy feedback in a physically
motivated and self-consistent way, we follow purely phenomenological recipes
to include non-gravitational gas heating, which are described in the
following. The total heating energy budget in such schemes was chosen so as
to agree with observations of SN products in the diffuse ICM (Finoguenov et
al. 2001a) and in stars (Renzini 1997).

\begin{table}[!t]
{
\footnotesize
{\renewcommand{\arraystretch}{0.9}\renewcommand{\tabcolsep}{0.09cm}
\caption{\footnotesize
Characteristics of the simulations
\label{t:run}}

\begin{tabular}{lcccl}
\hline
Simulation & \multicolumn{1}{c}{$f_{\rm gas}$} & {$f_{*}/f_{\rm bar}^{\dag}$}
& \multicolumn{1}{c}{E$^{\flat}$} & Line coding  \\
\hline 

\\
\multicolumn{5}{l}{Virgo; $~~M_{500}=23$;
$~~N_{\rm g}\simeq 1.5\times 10^5$; Fig.1(c)} \\
GH & 0.12 & ---  & ---  & dot-long dash  \\
SF & 0.09 & 0.25 & ---  & short-long dash  \\
S25-9 & 0.11 & 0.12 & 0.5 & long dash \\
S50-9 & 0.11 & 0.07 & 0.9 & dot-short dash \\
S50-3 & 0.10 & 0.18 & 0.8 & dotted \\
K75-3 & 0.10 & 0.18 & 0.75 & solid \\

\\
\multicolumn{5}{l}{Group 1; $~~M_{500}=3.8$;
$~~N_{\rm g}\simeq 1.8\times 10^5$; Fig.1(b)}\\
GH & 0.12 & --- & ---  & dot-long dash \\
SF & 0.08 & 0.37 & ---  & short-long dash \\
S25-9$^{\natural}$ & 0.11 & 0.11 & 0.4  & long dash \\
S50-9$^{\natural}$ & 0.11 & 0.02 & 0.9 & dot-short dash \\
S50-3 & 0.08 & 0.26 & 0.9 & dotted \\
K75-3 & 0.07 & 0.27 & 0.75 & solid \\

\\
\multicolumn{5}{l}{Group 2; $~~M_{500}=1.6$;
$~~N_{\rm g}\simeq 7.8\times 10^4$ thick lines in Fig.1(a)}\\
GH & 0.13 & --- & ---  & dot-long dash \\
SF & 0.08 & 0.37 & ---  & short-long dash  \\
S25-9$^{\natural}$ & 0.11 & 0.17 & 0.4  & long dash \\
S50-9$^{\natural}$ & 0.12 & 0.02 & 0.9 & dot-short dash \\
S50-3 & 0.08 & 0.27 & 0.9 & dotted \\
K75-3 & 0.07 & 0.28 & 0.75 & solid \\

\\
\multicolumn{5}{l}{Group 3; $~~M_{500}=1.4$;
$~~N_{\rm g}\simeq 7.1\times 10^4$ thin lines in Fig.1(a)}\\
GH & 0.13 & --- & ---  & dot-long dash \\
SF & 0.08 & 0.38 & ---  & short-long dash  \\
S25-9$^{\natural}$ & 0.11 & 0.04 & 0.4  & long dash \\
S50-9$^{\natural}$ & 0.10 & 0.00 & 0.8 & dot-short dash \\
S50-3 & 0.06 & 0.25 & 0.6 & dotted \\
K75-3 & 0.05 & 0.26 & 0.75 & solid \\
\hline
\end{tabular}
\begin{enumerate}
\item[$^{\dag}$]{\footnotesize ~ $f_{\rm bar}=0.13$, i.e. initial baryon fraction}
\item[$^{\flat}$]{\footnotesize ~ keV particle$^{-1}$, using  $E_h= (3/2) kT$}
\item[$^{\natural}$]{\footnotesize ~ low-resolution run}
\end{enumerate}
}}
\vspace*{-0.8cm}
\end{table}

\begin{figure*}
\includegraphics[width=8.cm]{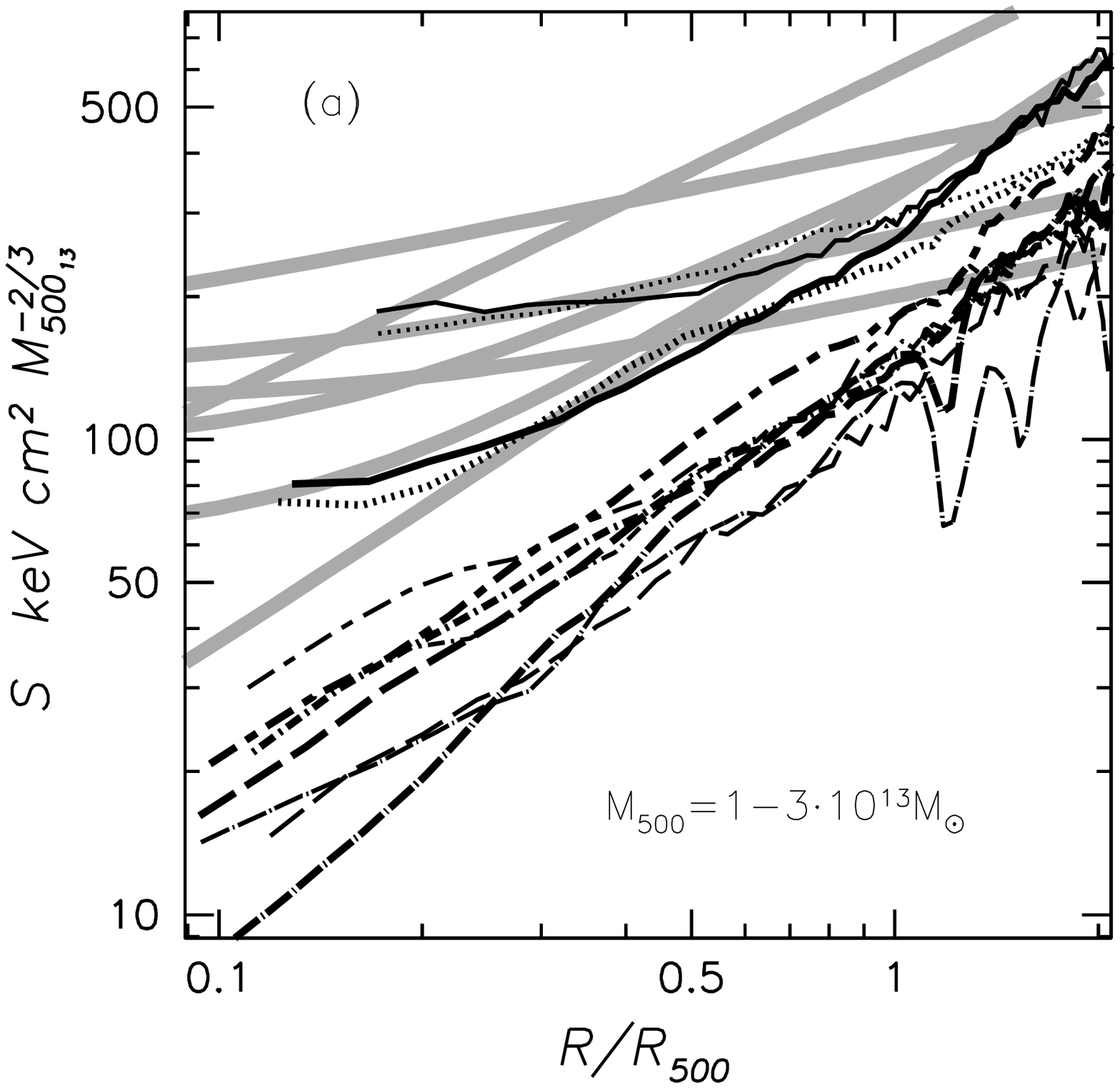}\hfill\includegraphics[width=8.cm]{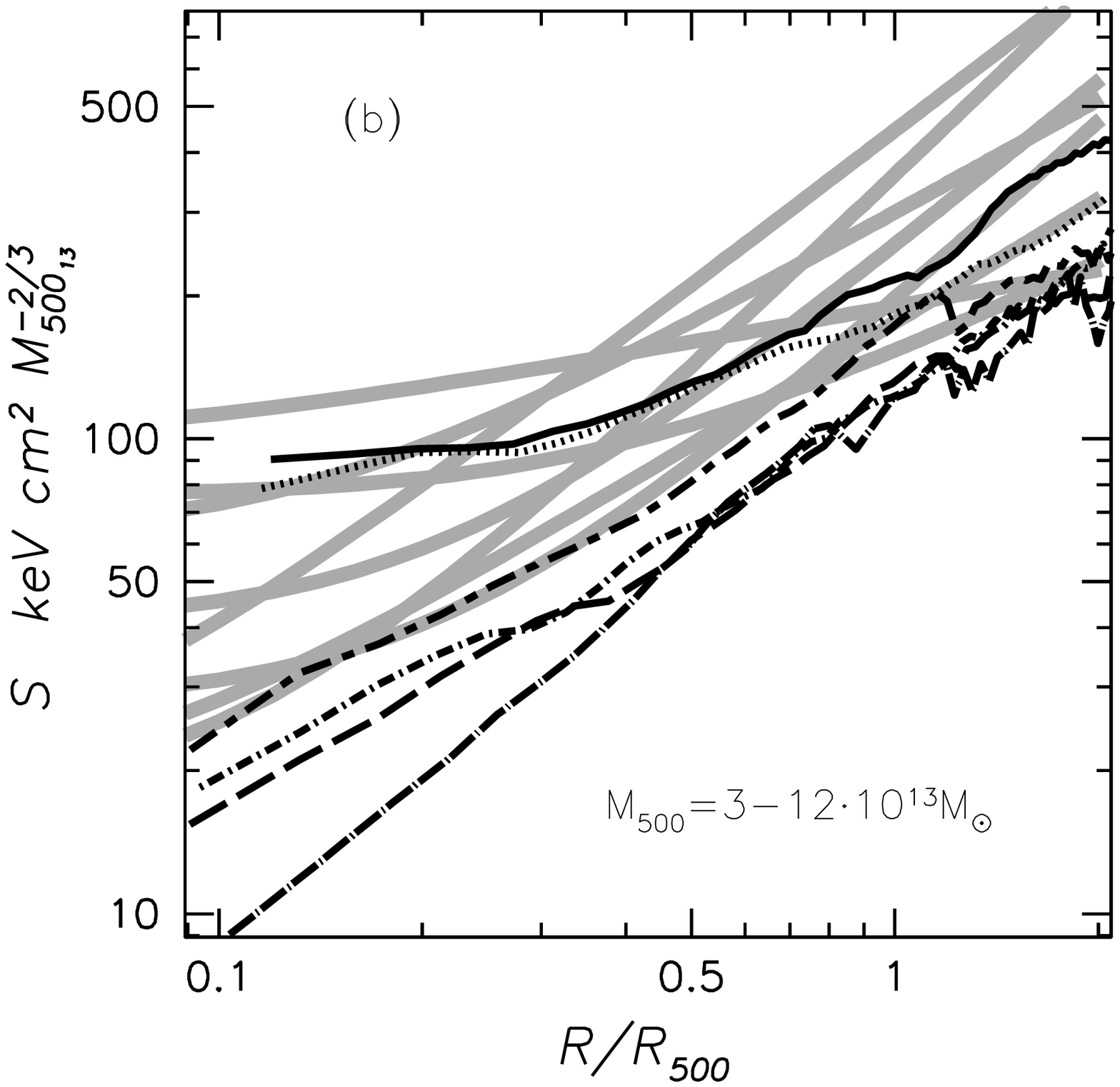}

\includegraphics[width=8.cm]{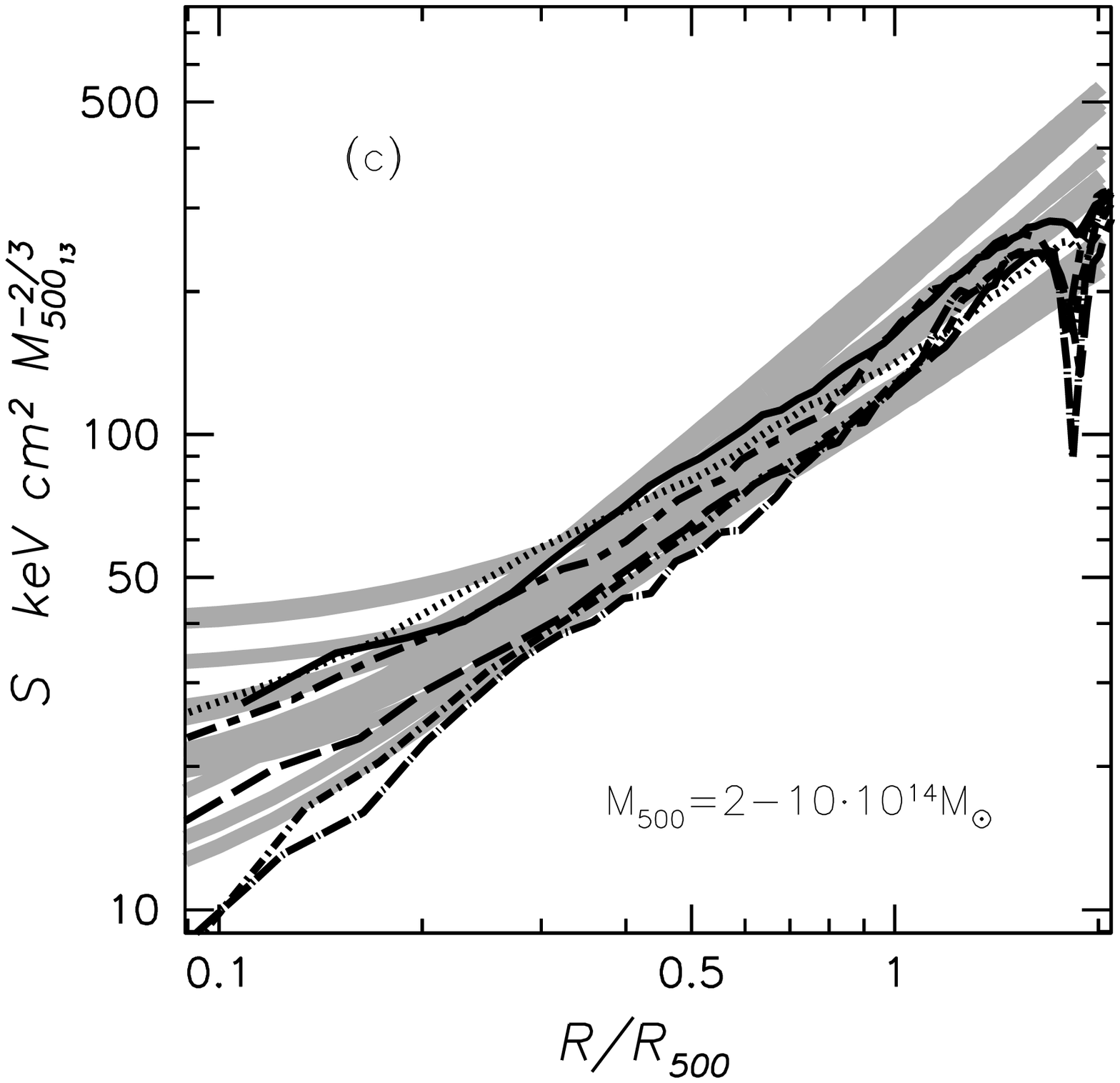}\hfill\includegraphics[width=8.cm]{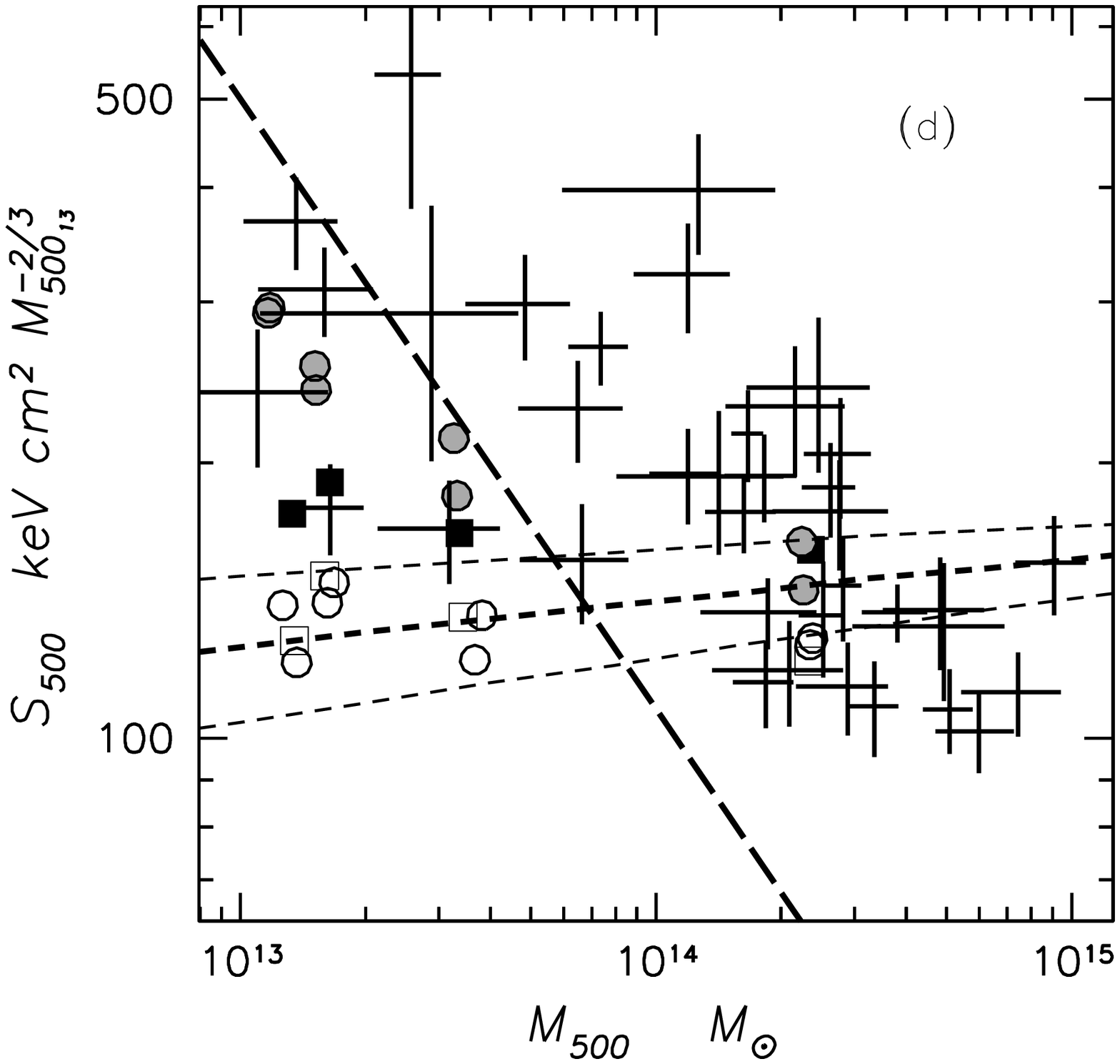}

\figcaption{Entropy profiles scaled by the total mass in units of $10^{13}$
\msun, measured at overdensity $\Delta_{500}$ and plotted against a scaling
radius measured as a fraction of $r_{500}$. Gray lines indicate the data.
Black lines denote various simulation runs: dot-long-dash for the GH runs,
short-long-dash for SF runs, long dash for the S25-9 runs, dot-short-dash
for the S50-9 runs, dotted for the S50-3 runs and solid for the K75-3 runs
(see Tab.\ref{t:run}). Comparison is made for the three mass ranges
separately in panels (a--c). Panel (d) shows a comparison at $r_{500}$,
where data points are shown as crosses, circles indicate simulations with
preheating, with filled circles representing the $z_h=3$. Open squares show
the purely gravitational runs and filed squares indicate the effect of
cooling. Dashed lines represent the prediction for shock heating (see
Appendix A in Finoguenov et al. 2002). The long-dashed line indicates the
effect of the preheating value of 500 keV cm$^2$.
\label{f:ent}
}
\end{figure*}
\vspace*{-0.2cm}
\begin{description}
\item[(a)] Imposing a minimum entropy floor, $S_{\rm floor}$, at some heating
redshift $z_h$ (e.g. Tozzi \& Norman 2001; Bialek et al. 2001; Borgani et
al. 2002). The entropy floor is created by increasing the temperature of
those gas particles having $S<S_{\rm floor}$, while leaving unchanged those
particles staying already at a higher adiabat. We choose here $z_h=9$, so as
to heat the diffuse gas well before a substantial amount of it can cool
within collapsed halos, and $z_h=3$, which is closer to the epoch at which
star formation peaks in the proto-cluster region. We realize three series of
runs, corresponding to $S_{\rm floor}=25$ and 50 keV cm$^2$ at $z_h=9$ (S25-9
and S50-9, respectively) and to the higher floor value at $z_h=3$ (S50-3)
(see fourth column of Table 1, for the average amount of heating received by
the gas particles ending up within the halo virial regions at $z=0$).

\item[(b)] Heating with an equal amount of energy per particle,
$E_h=0.75$ keV/particle, at $z_h=3$ (K75-3). This energy budget is
comparable to that provided by the S50-3 scheme.  Such a prescription
leads to an enhanced non-gravitational heating in the low density gas
and is characterized by a small fraction of subsequent radiative
losses of the feedback energy, 20\% compared to 80\% in the S50-3
run. These radiative losses when subtracted from the input energy can
be used as a measure of the type II SN feedback energy retained in the
gas. This effect should not be confused with the result of radiative
cooling to increase the mean specific energy of the remaining gas by
turning its colder component into stars.
\end{description}
\vspace*{-0.2cm} 
The simulations presented here neglect the influence of metal abundance on the
cooling function, which correspond to the conditions in the pristine
gas. While this is an oversimplification for the low-redshift ICM, it is
worth wondering {\it how} the observed $\simeq 0.3\,Z_\odot$ can be
reproduced. Pipino et al. (2002) considered an iron enrichment on the
precollapse ratio of stars and gas, and concluded that only 0.1 Fe$_\odot$
abundance could be reached with a Salpeter IMF, corresponding to a mild
underestimation of cooling in our simulations (a factor of 1.5 for the gas
at temperature of 1 keV). Using the ratio between iron mass and stellar
light Finoguenov et al. (2000) showed that high Fe abundance observed in
groups and cluster cores is a result of a reduced gas fraction. Thus, before
assuming an observed Fe abundance for the cooling function, one has to
produce a strong deviation from self-similarity with low-metallicity
cooling, as we do in this {\it Letter}. A subsequent increase of Fe
abundance will further rise the minimum entropy level required for gas
particles to remain in the hot phase.

The results from simulations are compared with the ASCA observations for a
sample of 38 groups and clusters of galaxies, presented in Finoguenov et
al. (2000, 2001a, 2002).  This large sample has been assembled during ten
years of telescope operation, with large spatial coverage for individual
systems, possible due to low detector background and an advanced stage of
the instrument calibration. The observations provide entropy, total
gravitational mass, and heavy element profiles. To compare with simulations
we will use entropy profiles scaled by $M_{500}^{2/3}$. Under the assumption
of hydrostatic equilibrium, $T\propto M^{2/3}$, so that the above entropy
rescaling produces overlapping profiles for different mass objects, as long
as the ICM behaves in a self-similar way.  As for the observed fraction of
cluster baryons locked into stars, $f_*$, Balogh et al. (2001) estimated it
to be 5 (or 10) \% from the 2MASS K-band luminosity function by Cole et
al. (2001), assuming a Kennicutt (1983) (or Salpeter, 1955) initial mass
function (IMF). 
Also, it is clear that this will represent a convergent
estimate as long as the galaxy survey is deep enough to trace most part of
the K-band luminosity. 
Quite recently, Huang et al. (2002) used the Hawaii--AAO survey to estimate
the K-band galaxy luminosity function. Although this survey covers a small
area, it is much deeper than the 2MASS. As a result these authors found the
K-band luminosity density to be a factor two higher than that derived from
the 2MASS, which would roughly imply a proportionally higher value for
$f_*$. Furthermore, it is also questionable whether the value for the field
could be taken as representative for clusters. Owing to such uncertainties,
we adopt the $f_*$ to range in the 10--20 per cent interval, as
representative of what is currently indicated by data.  Differences in the
fraction of diffuse gas, $f_{\rm gas}$, as reported in Col. 2 of Table
\ref{t:run}, are both due to different efficiency of star formation and to
the different spatial extent of the gas associated to the different heating
schemes.  To provide an unbiased comparison for the stellar fraction, we
compare it to the initial baryon mass, by using the total mass and the
initial baryon fraction.

\section{Results and discussion}
Figure \ref{f:ent} summarizes the comparison between the entropy properties
of the ICM for simulated and observed galaxy systems. Panel (a) shows the
entropy profiles for observed groups with $M_{500}=(1-3)\times
10^{13}M_{\odot}$, compared to the entropy profiles from the simulations of
Group-2 and Group-3, panel (b) is for data on groups and poor clusters with
$M_{500}=(3-12)\times 10^{13}M_{\odot}$ and the simulation of the Group-1,
panel (c) is for observed clusters with $M_{500}=(2-10)\times
10^{14}M_{\odot}$ and the ``Virgo'' cluster simulation. Panel (d) shows the
positions of real and simulated structures on the $(S_{500},M_{500})$ plane
(Finoguenov et al. 2002; here $S_{500}$ is the entropy computed at
$r_{500}$). Entropy profiles from simulations are plotted down to the radius
which contains 100 gas particles. This criterion has been shown by Borgani
et al. (2002) to avoid numerical effects and provides results which are
stable against numerical resolution.

In panel (a), where we realize the best matching between the masses of
simulated and observed structures, the preheating runs with $z_h=3$
reproduce all the entropy characteristics of the observational data.  We
note that the S50-3 and K75-3 runs are characterized by comparable amounts
of both extra energy and resulting stellar fraction, despite the fact that
the two heating recipes spread a different amount of energy to gas at
different density.  The differences between the two prescriptions come at
radii exceeding $0.5R_{500}$, where the S50-3 approaches the entropy level
produced by the gravitational heating case, while in the K75-3 run the
entropy keeps increasing parallel to the GH simulations. This behavior could
be attributed to the way the gas has been distributed at $z_h=3$, before its
accretion onto clusters.  Assigning an equal amount of extra heating to all
gas particles is somewhat equivalent to the effect produced by an accretion
shock of a fixed strength. Therefore, if self-similarity of the gas holds,
the effect of heating will result in a radial behavior of the entropy, which
is similar to that of gravitational heating. On the contrary, the S50-3
heating has a marginal effect in increasing the temperature of low-density
regions and, therefore, is less efficient in preventing the occurrence of
the gravitational shock. As a result, gas undergoes roughly the same amount
of gravitational heating as in the self-similar case, whose entropy level is
in fact attained in the outer halo regions.  At radii lower than
$0.5$--$1\,R_{500}$ in Fig.1(a-b) the effect of heating is the same in both
the K75-3 and the S50-3 runs. According to the energy distribution, one
expects the S50-3 to achieve a higher entropy at lower radii. Yet the
profiles are nearly the same. This points to an important role of
cooling in equating the K75-3 and the S50-3 runs at the center. The
difference between the two runs is in the cooling times at high-density
regions, which are shorter for K75-3. A similarity in both the entropy
profiles and cold fractions indicate that all the high-density gas has been
cooled irregardless of the amount of energy put there.

Cooling has been suggested as a mechanism to increase the observed entropy
level of the ICM (e.g. Voit et al. 2002, and references therein). Although,
by their nature, radiative losses decrease gas entropy, low-entropy
particles, which have short cooling time, $t_{\rm cool}$, are removed from
the X-ray emitting phase, thus leaving only high entropy gas, which has
longer $t_{\rm cool}$. However, our results actually show that including
star formation and no extra heating produce too steep entropy profiles, an
effect which is more evident in the runs of groups. This result suggests
that a fair population of low-entropy particles is still left in the diffuse
phase during the cooling process. Since cooling of the preheated component
is negligible, a combination of preheating and cooling better matches the
conditions needed for the model of Voit \& Bryan (2001) to produce a plateau
in the entropy profile: a required separation between the low-entropy gas,
which is able to cool and the high entropy gas retained in the hot phase.
At outskirts of the groups, however, as shown in Fig.1(d), the observed
entropy could only be explained by runs with preheating at $z=3$.

In Fig.\ref{f:fgas} we outline the difference between simulation runs
in accounting at the same time for gas entropy and star fraction.  As
expected, the effect of including only gas cooling and star formation
is an over-production of the amount of gas locked in the cold phase,
while it is only marginal accounting for the observed gas entropy level.
The preheating at $z_h=9$ results in a much delayed star-formation,
pointing out the need of strongly reducing the energetics of
preheating at that epoch. However, even with the amount of energy
assigned in our runs, it is not possible to reproduce the high
entropy level observed in groups.

\vspace*{-0.05cm} \includegraphics[width=8.cm]{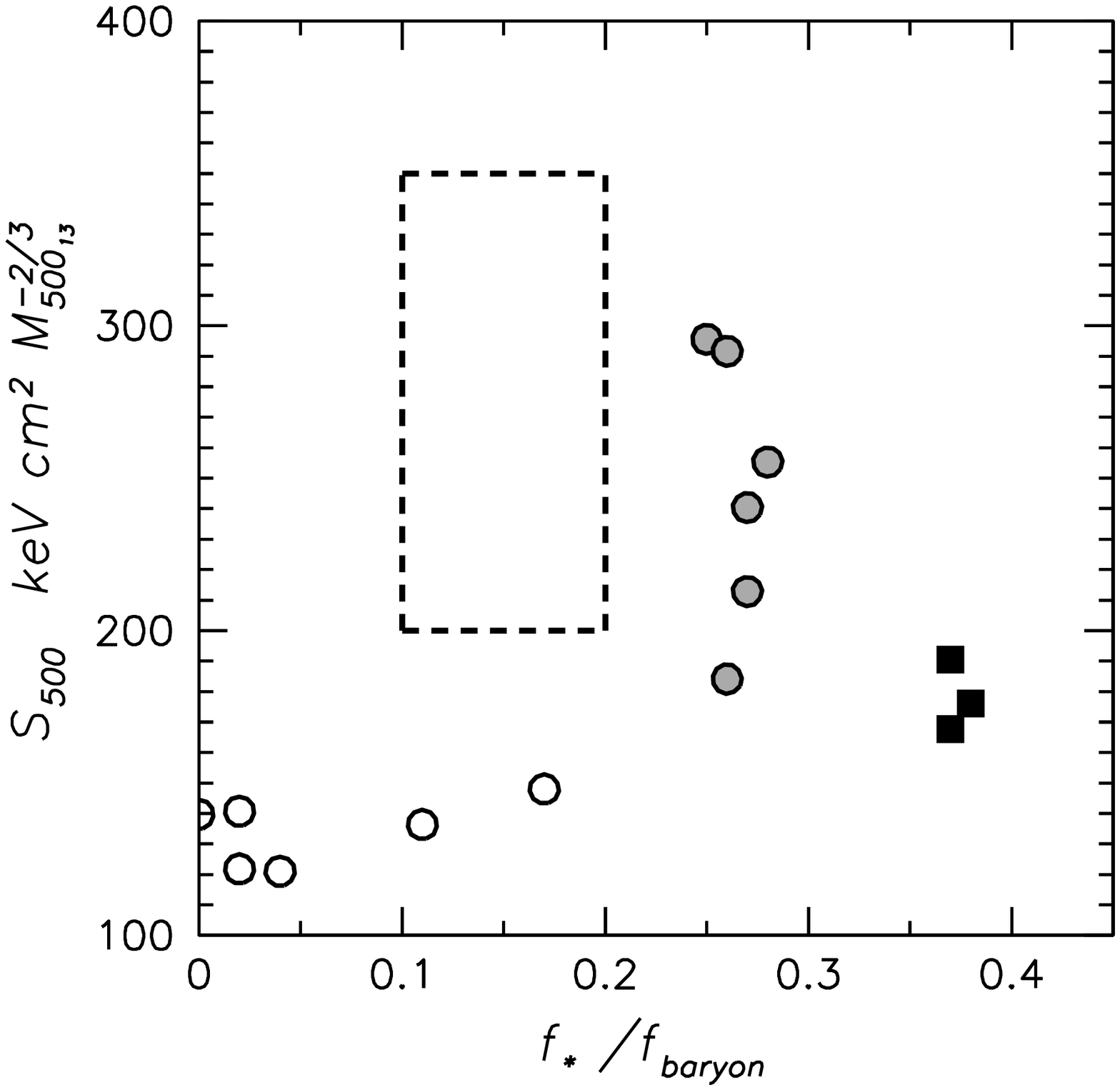}
\figcaption{Simultaneous reproduction of both entropy and amount of
star-formation in simulations of clusters and groups. Filled squares are for
the runs only including cooling and star formation. Open (filled) circles
denote the simulations with preheating at $z_h=9$ ($z_h=3$). The box denotes
observational limits discussed in the text.
\label{f:fgas}
}
\vspace*{-0.05cm}

Heating at $z_h=3$ performs better in reproducing the observed entropy
level, while it still results in some overproduction of the stellar
mass fraction. Besides the uncertainties in the observational
determination of $f_*$, there is also a number of physical effects,
which are not included in our simulations and which should help in
further alleviating gas overcooling. For instance, our preheated
runs with $z_h=3$ ignore a possible additional feedback from the
formation of the first stars (e.g. Madau, Ferrara \& Rees 2001),
which, according to our experience with the $z_h=9$ run, could delay
accretion of the gas on the forming halos, both delaying the
star-formation (Tornatore et al. 2002) and, in
combination with the $z_h=3$ preheating, reducing the fraction of the
accreted gas, which could later be available for the star-formation.
As a concluding remark, one of the most important results of our analysis is
that the entropy profiles in simulations do not depend only on the amount of
extra energy, but also on the details of energy ejection, such as the epoch
and the spatial distribution. We have demonstrated that the processes of gas
accretion, recorded in the gas entropy profiles, also regulate the
star-formation. 

\smallskip

{\it Acknowledgments.} We are very grateful to Volker Springel for providing
support in using GADGET, the referee, Mark Voit, for constructive comments,
and Trevor Ponman and Paolo Tozzi for useful discussions. The simulations
have been run on the IBM-SP3 at the Computing Centers of the Astronomical
Observatory in Catania and of the University of Trieste, and on the IBM-SP4
at CINECA in Bologna, under INAF-CINECA grant.

\vspace*{-0.5cm}

\bibliographystyle{aabib99}

\end{document}